\documentclass[dvips]{article}
\usepackage{supertabular,lscape,epsfig}
\usepackage{amssymb}
\usepackage{amsmath}
\DeclareSymbolFont{ppa}{OT1}{ppl}{m}{it}
\DeclareMathSymbol{\vv}{\mathalpha}{ppa}{'166}

\DeclareSymbolFont{ppa}{OT1}{ppl}{m}{it}
\DeclareMathSymbol{\vv}{\mathalpha}{ppa}{'166}

\begin{document}

\newcommand{\dd}{\,{\rm d}}
\newcommand{\ie}{{\it i.e.},\,}
\newcommand{\etal}{{\it et al.\ }}
\newcommand{\eg}{{\it e.g.},\,}
\newcommand{\cf}{{\it cf.\ }}
\newcommand{\vs}{{\it vs.\ }}
\newcommand{\zdot}{\makebox[0pt][l]{.}}
\newcommand{\up}[1]{\ifmmode^{\rm #1}\else$^{\rm #1}$\fi}
\newcommand{\dn}[1]{\ifmmode_{\rm #1}\else$_{\rm #1}$\fi}
\newcommand{\upd}{\up{d}}
\newcommand{\uph}{\up{h}}
\newcommand{\upm}{\up{m}}
\newcommand{\ups}{\up{s}}
\newcommand{\arcd}{\ifmmode^{\circ}\else$^{\circ}$\fi}
\newcommand{\arcm}{\ifmmode{'}\else$'$\fi}
\newcommand{\arcs}{\ifmmode{''}\else$''$\fi}
\newcommand{\MS}{{\rm M}\ifmmode_{\odot}\else$_{\odot}$\fi}
\newcommand{\RS}{{\rm R}\ifmmode_{\odot}\else$_{\odot}$\fi}
\newcommand{\LS}{{\rm L}\ifmmode_{\odot}\else$_{\odot}$\fi}

\newcommand{\Abstract}[2]{{\footnotesize\begin{center}ABSTRACT\end{center}
\vspace{1mm}\par#1\par
\noindent
{~}{\it #2}}}

\newcommand{\TabCap}[2]{\begin{center}\parbox[t]{#1}{\begin{center}
  \small {\spaceskip 2pt plus 1pt minus 1pt T a b l e}   
  \refstepcounter{table}\thetable \\[2mm]
  \footnotesize #2 \end{center}}\end{center}}

\newcommand{\TableSep}[2]{\begin{table}[p]\vspace{#1}
\TabCap{#2}\end{table}}

\newcommand{\FigCap}[1]{\footnotesize\par\noindent Fig.\  %
  \refstepcounter{figure}\thefigure. #1\par}

\newcommand{\TableFont}{\footnotesize}
\newcommand{\TableFontIt}{\ttit}
\newcommand{\SetTableFont}[1]{\renewcommand{\TableFont}{#1}}

\newcommand{\MakeTable}[4]{\begin{table}[p]\TabCap{#2}{#3}
  \begin{center} \TableFont \begin{tabular}{#1} #4 
  \end{tabular}\end{center}\end{table}}

\newcommand{\MakeTableSep}[4]{\begin{table}[p]\TabCap{#2}{#3}
  \begin{center} \TableFont \begin{tabular}{#1} #4
  \end{tabular}\end{center}\end{table}}
\newcommand{\TabCapp}[2]{\begin{center}\parbox[t]{#1}{\centerline{
  \small {\spaceskip 2pt plus 1pt minus 1pt T a b l e}
  \refstepcounter{table}\thetable}
  \vskip2mm
  \centerline{\footnotesize #2}}
  \vskip3mm
\end{center}}

\newcommand{\MakeTableSepp}[4]{\begin{table}[p]\TabCapp{#2}{#3}\vspace*{-.7cm}
  \begin{center} \TableFont \begin{tabular}{#1} #4 
  \end{tabular}\end{center}\end{table}}

\newfont{\bb}{ptmbi8t at 12pt}
\newfont{\bbb}{cmbxti10}
\newfont{\bbbb}{cmbxti10 at 9pt}
\newcommand{\uprule}{\rule{0pt}{2.5ex}}
\newcommand{\douprule}{\rule[-2ex]{0pt}{4.5ex}}
\newcommand{\dorule}{\rule[-2ex]{0pt}{2ex}}
\def\thefootnote{\fnsymbol{footnote}}

\newenvironment{references}%
{
\footnotesize \frenchspacing
\renewcommand{\thesection}{}
\renewcommand{\in}{{\rm in }}
\renewcommand{\AA}{Astron.\ Astrophys.}
\newcommand{\AAS}{Astron.~Astrophys.~Suppl.~Ser.}
\newcommand{\ApJ}{Astrophys.\ J.}
\newcommand{\ApJS}{Astrophys.\ J.~Suppl.~Ser.}
\newcommand{\ApJL}{Astrophys.\ J.~Letters}
\newcommand{\AJ}{Astron.\ J.}
\newcommand{\IBVS}{IBVS}
\newcommand{\PASP}{P.A.S.P.}
\newcommand{\Acta}{Acta Astron.}
\newcommand{\MNRAS}{MNRAS}
\renewcommand{\and}{{\rm and }}
\section{{\rm REFERENCES}}
\sloppy \hyphenpenalty10000
\begin{list}{}{\leftmargin1cm\listparindent-1cm
\itemindent\listparindent\parsep0pt\itemsep0pt}}%
{\end{list}\vspace{2mm}}

\def\TYLDA{~}
\newlength{\DW}
\settowidth{\DW}{0}
\newcommand{\dw}{\hspace{\DW}}

\newcommand{\refitem}[5]{\item[]{#1} #2%
\def\REFARG{#3}\ifx\REFARG\TYLDA\else, {\it#3}\fi
\def\REFARG{#4}\ifx\REFARG\TYLDA\else, {\bf#4}\fi
\def\REFARG{#5}\ifx\REFARG\TYLDA\else, {#5}\fi.}

\newcommand{\Section}[1]{\section{\hskip-6mm.\hskip3mm#1}}
\newcommand{\Subsection}[1]{\subsection{#1}}
\newcommand{\Acknow}[1]{\par\vspace{5mm}{\bf Acknowledgements.} #1}
\pagestyle{myheadings}

\newcommand{\xrule}{\rule{0pt}{2.5ex}}
\newcommand{\xxrule}{\rule[-1.8ex]{0pt}{4.5ex}}
\def\thefootnote{\fnsymbol{footnote}}
\begin{center}
{\Large\bf The Optical Gravitational Lensing Experiment.\\
\vskip3pt
Ellipsoidal Variability of Red Giants\\
\vskip6pt
in the Large Magellanic Cloud\footnote{Based on observations obtained with the
1.3~m Warsaw telescope at the Las Campanas Observatory of the Carnegie
Institution of Washington.}}
\vskip1.2cm
{\bf I.~~S~o~s~z~y~\'n~s~k~i$^{1,2}$,~~A.~~U~d~a~l~s~k~i$^1$,~~M.~~K~u~b~i~a~k$^1$,\\
M.\,K.~~S~z~y~m~a~{\'n}~s~k~i$^1$,~~G.~~P~i~e~t~r~z~y~\'n~s~k~i$^{1,2}$,~~K.~~\.Z~e~b~r~u~\'n$^1$,\\
O.~~S~z~e~w~c~z~y~k$^1$,~~\L.~~W~y~r~z~y~k~o~w~s~k~i$^1$\\
and~~W.\,A.~~D~z~i~e~m~b~o~w~s~k~i$^1$}
\vskip8mm
{$^1$Warsaw University Observatory, Al.~Ujazdowskie~4, 00-478~Warszawa, Poland\\
e-mail: \small{(soszynsk,udalski,mk,msz,pietrzyn,zebrun,szewczyk,wyrzykow,wd)@astrouw.edu.pl}\\
$^2$ Universidad de Concepci{\'o}n, Departamento de Fisica, Casilla 160--C, Concepci{\'o}n, Chile}
\end{center}

\vskip1.6cm

\Abstract{We used the OGLE-II and OGLE-III photometry of red giants in the
Large Magellanic Cloud to select and study objects revealing ellipsoidal
variability. We detected 1546 candidates for long period ellipsoidal
variables and 121 eclipsing binary systems with clear ellipsoidal
modulation. The ellipsoidal red giants follow a period--luminosity ({\it PL})
relationship (sequence~E), and the scatter of the relation is correlated
with the amplitude of variability: the larger the amplitude, the smaller
the scatter.

We note that some of the ellipsoidal candidates exhibit simultaneously OGLE
Small Amplitude Red Giants pulsations. Thus, in some cases the Long
Secondary Period (LSP) phenomenon can be explained by the ellipsoidal
modulation.

We also select about 1600 red giants with distinct LSP, which are not
ellipsoidal variables. We discover that besides the sequence D in the {\it
PL} diagram known before, the LSP giants form additional less numerous
sequence for longer periods. We notice that the {\it PL} sequence of the
ellipsoidal candidates is a direct continuation of the LSP sequence toward
fainter stars, what might suggest that the LSP phenomenon is related to
binarity but there are strong arguments against such a possibility.

About 10\% of the presented light curves reveal clear deformation by the
eccentricity of the system orbits. The largest estimated eccentricity in
our sample is about~0.4.

All presented data, including individual {\it BVI} observations and finding
charts are available from the OGLE Internet archive.}{binaries: close --
binaries: eclipsing -- Stars: late-type -- Magellanic Clouds}

\Section{Introduction}
Ellipsoidal variability is a phenomenon often observed in close binary
systems. It is caused by aspect changes of the component deviated from
spherical symmetry by tidal interactions with its companion. A typical
ellipsoidal light curve is sinusoidal-shaped, with two maxima and two
minima per orbital period. Light variations should peak at orbital phases
0.25 and 0.75, when maximal area of the Roche-lobe is projected. In general
the depths of the minima are not equal, which is the result of next order
effects of the tidal distortion.

Ellipsoidal variables are relatively rarely studied objects, and long
period ellipsoidal variables are almost unknown. Morris (1985) gathered all
ellipsoidal variables known at that time, and listed 20 confirmed and 20
suspected objects of this type. Only one variable star from this catalog
(T~CrB) had period longer than 100~days. Since then, the number of known
ellipsoidal variables has increased, but still, the list of confirmed
ellipsoidal red giants is very short. However, the analysis of the
ellipsoidal variations is a useful tool for studying various types of close
binary systems: X-ray binaries, cataclysmic binaries, symbiotic stars,
early type near contact systems.

In recent years gravitational microlensing surveys have revolutionized our
kno\-wledge about variable red giants. Wood \etal (1999) presented
period--luminosity ({\it PL}) diagram of the long period variables which
showed series of distinct parallel sequences, marked with letters A--E.

In the previous paper (Soszy{\'n}ski \etal 2004), we selected and studied
about 15\,000 red giants occupying most often the sequence~A. We showed
that these objects, named by Wray, Eyer and Paczy{\'n}ski (2004) OGLE Small
Amplitude Red Giants (OSARG), constitute a separate type of pulsating
giants, different than ``classical'' semi-regular variables (SRV).

In this paper we present a sample of variable red giants constituting {\it
PL} sequence~E, spreading below the sequences A--D. Wood \etal (1999)
suggested that most of these stars are contact binaries. We studied
photometry of these objects and noticed that, indeed, the majority of the
light curves are probably ellipsoidal or eclipsing close binary systems.
Our sample is, then, a natural extension of the catalog of eclipsing binary
systems in the LMC prepared by Wyrzykowski \etal (2003).

\Section{Observations and Data Reductions}
Observations presented in this paper were carried out with the 1.3-m Warsaw
telescope at the Las Campanas Observatory, Chile, operated by the Carnegie
Institution of Washington. Central parts of the LMC were added to the list
of regularly monitored fields in January 1997, \ie at the beginning of the
second phase of the OGLE project (OGLE-II). The telescope was then equipped
with the ``first generation'' camera with the SITe ${2048\times2048}$ CCD
detector. The pixel size was 24~$\mu$m resulting in 0.417~arcsec/pixel
scale. Observations of the LMC were performed in the ``slow'' reading
mode of the CCD detector with the gain 3.8~e$^-$/ADU and readout noise of
about 5.4~e$^-$. Details of the instrumentation setup can be found in
Udalski, Kubiak and Szyma{\'n}ski (1997).

In June 2001 the third stage of the OGLE experiment (OGLE-III) began. The
Warsaw telescope was equipped with a ``second generation'' CCD mosaic
camera consisting of eight SITe ST-002a CCD detectors with
${2048\times4096}$ pixels of 15~$\mu$m size (Udalski 2003). This
corresponds to 0.26~arcsec/pixel scale and the field of view of the whole
mosaic ${35\arcm\times35\arcm}$. The last observations presented in this
paper were collected in May 2004.

The vast majority of the observing points (430--860, depending on the
field) were obtained through the {\it I} filter, while in the {\it V} and
{\it B}-bands several dozen measurements were collected. The OGLE {\it BVI}
filters closely resemble the standard system. OGLE-II {\it I}-band
photometry was obtained using the Difference Image Analysis (DIA) method --
image subtraction algorithm developed by Alard and Lupton (1998) and Alard
(2000), and implemented by Wo{\'z}niak (2000). OGLE-III magnitudes come
from the standard data pipeline (Udalski 2003). {\it V} and {\it B}-band
photometry was performed with the modified version of the {\sc DoPhot}
package (Schechter, Mateo and Saha 1993). OGLE-II and OGLE-III data were
tied in the identical manner as in Soszy\'nski \etal (2004).

\Section{Selection of the Ellipsoidal Candidates}
We performed the period analysis for every star brighter than
${I=18}$~mag. We used {\it I}-band light curves, for which the majority of
observations had been collected and ran {\sc Fnpeaks} program
(Ko{\l}aczkowski 2003, private communication) to derive the most
significant periodicities.

For further analysis we selected stars with significant periodic light
variations, \ie objects for which the ratio of the highest peak in the
periodogram to the mean value was higher than~8. For this sample we
constructed the $\log{P}$--$W_I$ diagram, where $W_I$ is reddening free
Wesenheit index, defined as:
$$W_I=I-1.55(V-I)$$ 
where {\it I} and {\it V} are intensity mean magnitudes and 1.55 is the
mean ratio of total-to-selective absorption ${(A_I/E(V-I))}$. It appears that
$\log{P}$--$W_I$ diagrams can successfully replace period -- near-infrared
magnitude diagram, commonly used in the variable red giants studies.

Our $\log{P}$--$W_I$ diagram shows a series of sequences, presented for the
first time by Wood \etal (1999), and marked by letters A--E. In this paper
we analyze the sequence E, spreading below the ridges of OSARGs, SRV, Miras
and Long Secondary Periods (LSP), and extending down to our limiting
luminosity (18~mag).

\begin{figure}[p]
\centerline{\includegraphics{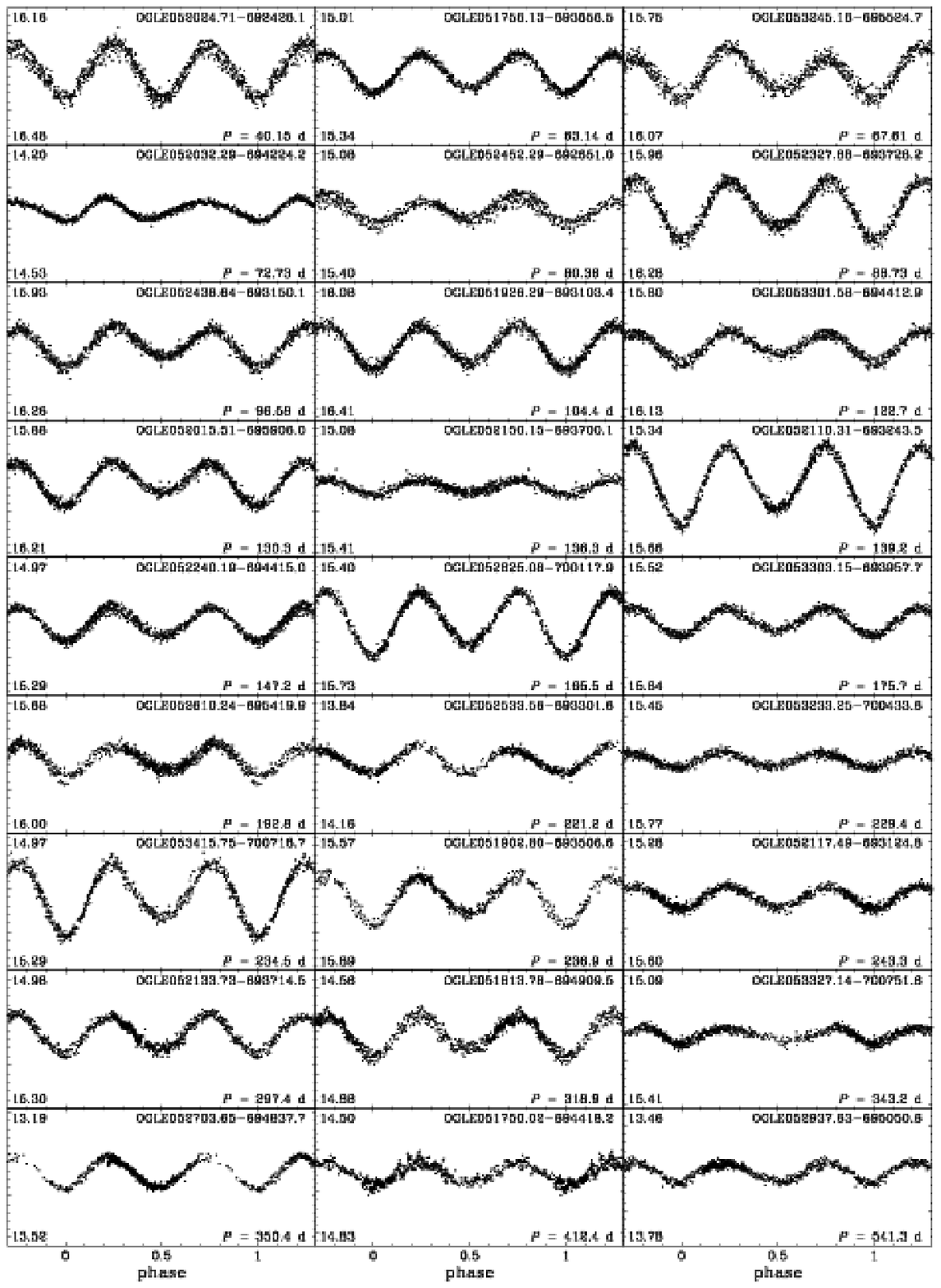}}
\FigCap{Light curves of the exemplary ellipsoidal red giants in the LMC.}
\end{figure} 

We selected and visually inspected the light curves of stars forming the
sequence~E. It appears that most of these objects are characterized by
nearly sinusoidal light curves with an amplitude between a few hundredths
and a few tenths of magnitude. Fig.~1 presents exemplary light curves of
these objects. It also appears that the majority of the light curves shows
alternately shallower and deeper minima, so the formal period is two times
longer than the one obtained automatically in our period search.

\begin{figure}[t]
\centerline{\includegraphics{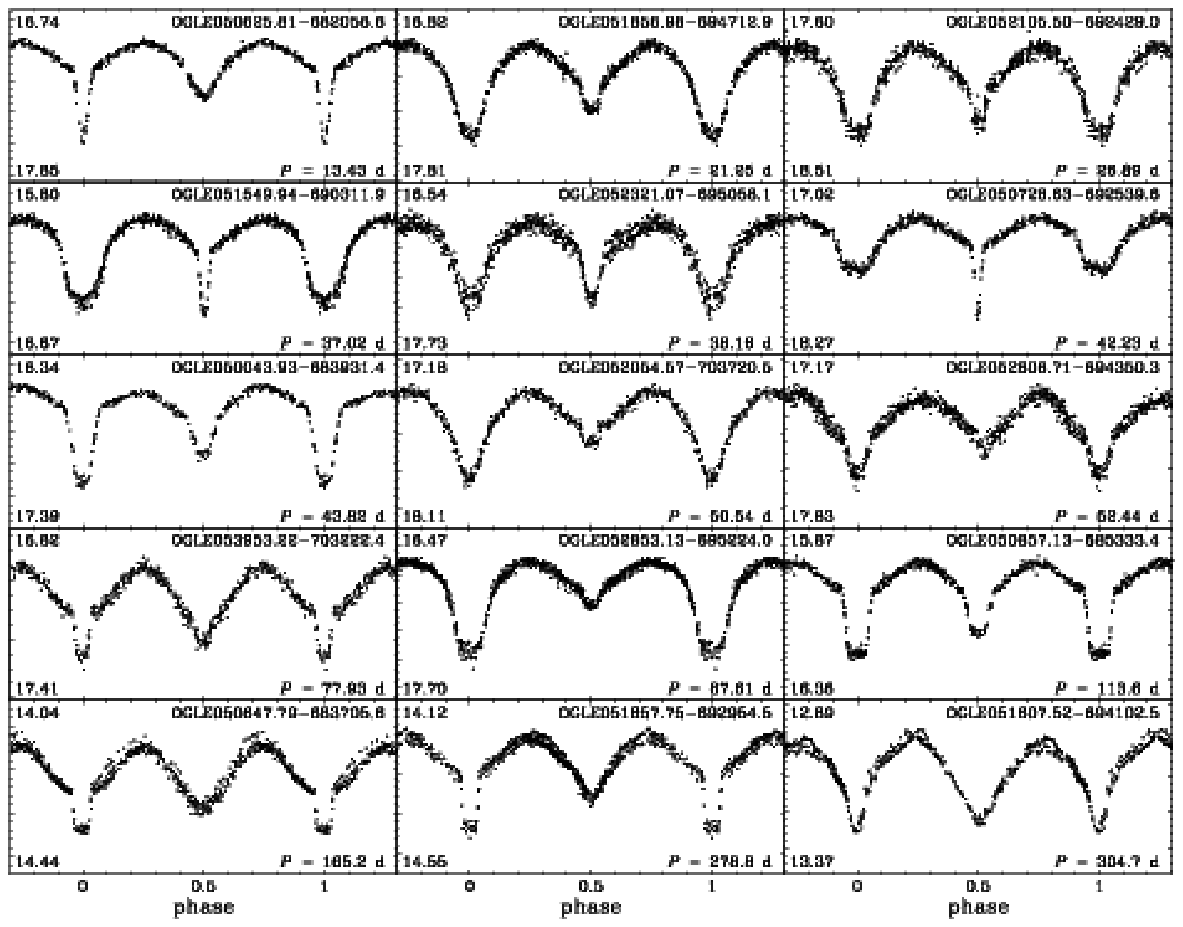}} 
\FigCap{Light curves of the exemplary eclipsing red giants in the LMC.}
\end{figure} 
Among selected objects we also found 121 eclipsing variables, most often
showing clear ellipsoidal effect. Several typical light curves showing
eclipses are presented in Fig.~2. We compared the eclipsing binaries with
the sample of sinusoidal variables, and found that when neglecting
eclipses, the shapes of the light curves from both samples are very
similar.

Taking into account this similarity of the eclipsing and ``sinusoidal''
light curves, as well as the same position in the color--luminosity and
period--luminosity diagrams, we interpret the ``sinusoidal'' objects as
binary systems with orbital inclinations too small to cause the
eclipses. Apparent variability of these objects is caused by the
ellipsoidal distortion of red giants.

To confirm whether a star is or is not an ellipsoidal variable, one needs
to obtain a radial velocity curve. We believe that the variability of the
vast majority of selected stars are caused by ellipsoidal modulation, but
one should remember that our sample may also contain a number of other
types of variables. We expect such contamination especially among stars
with the smallest amplitudes.

It is worth mentioning that ellipsoidal light curves which, on first sight,
do not show eclipses, can actually be eclipsing variables. First, there
might be grazing eclipses, which are not detectable by visual
inspection. Second, even complete eclipses can be very shallow when the
radius of the secondary component is significantly smaller than the radius
of the red giant. Some of these objects can be X-ray eclipsing systems,
which do not show eclipses in visual pass-bands, but they are observable in
X-rays.

\Section{Ellipsoidal and Eclipsing Variables}
We selected 1546 candidates for the ellipsoidal and 121 eclipsing 
variables in the LMC, with periods ranging from 15 to 600 days. Table~1 
lists first 70~candidates for the ellipsoidal variables from the field 
LMC\_SC1. The following columns present star ID, star number (consistent
with the LMC photometric maps, Udalski \etal 2000), equatorial
coordinates, RA and DEC, for the epoch 2000.0, periods in days
(referred to the double-peak light curves), moment of the zero phase
(corresponding to the deeper minimum), intensity mean {\it IVB} 
magnitudes, larger and smaller amplitudes of the {\it I}-band variability 
and remarks.

The list of all variables is available in the electronic form from the 
OGLE {\sc Internet} archive:

\begin{center}
{\it http://ogle.astrouw.edu.pl/} \\
{\it ftp://ftp.astrouw.edu.pl/ogle/ogle2/var\_stars/lmc/ell/}\\
\end{center}
or its US mirror
\begin{center}
{\it http://bulge.princeton.edu/\~{}ogle/}\\
{\it ftp://bulge.princeton.edu/ogle/ogle2/var\_stars/lmc/ell/}\\
\end{center}

Individual {\it BVI} measurements of all objects and finding charts are
also available from the OGLE {\sc Internet} archive. The lists contain
together 1753 entries but only 1667 objects, because 86 stars were
detected twice -- in the overlapping regions of adjacent fields. We
decided not to remove twice-detected stars from the final list, because
their measurements are independent in both fields and can be used for
testing quality of the data and the completeness of the sample.

\begin{figure}[!tb]
\vspace{-.7cm}
\centerline{\includegraphics[width=11.9cm]{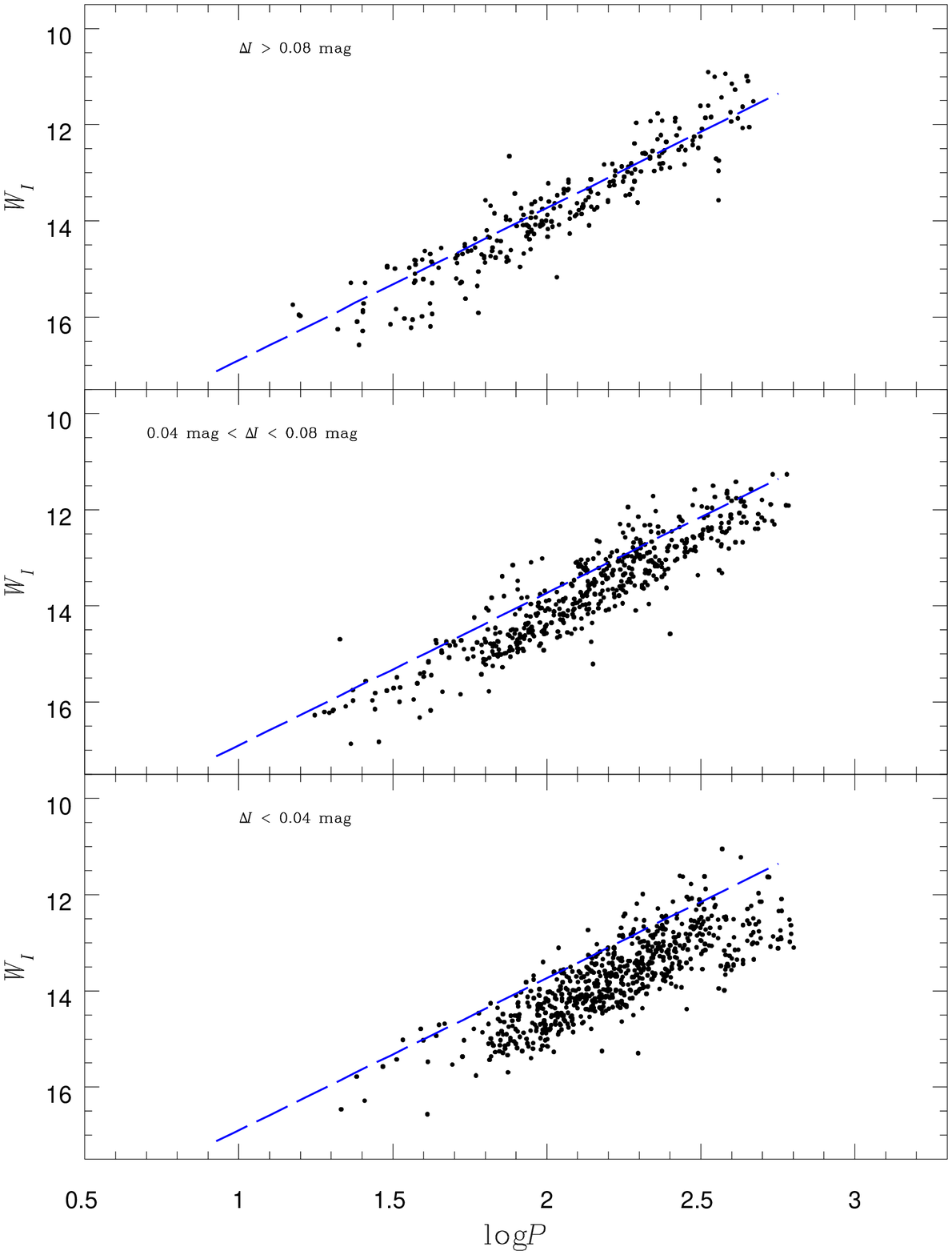}} 
\vspace{-0.3cm}
\FigCap{Period--$W_I$ diagrams for giants classified as ellipsodal
variables. {\it Upper panel} shows variables with the largest amplitudes
($\Delta I>0.08$~mag), {\it middle panel} presents variables with
intermediate amplitudes ($0.04<\Delta I<0.08$~mag), while {\it lower panel}
stars with the smallest amplitudes ($\Delta I<0.04$~mag).}
\end{figure} 

In Fig.~3 we plotted the $\log{P}$--$W_I$ diagrams for the ellipsoidal
binaries. The upper panel shows variables with the largest amplitudes
(${\Delta I>0.08}$~mag),\linebreak the middle panel presents variables with
intermediate amplitudes\linebreak (${0.04<\Delta I<0.08}$~mag), while the
lower panel -- stars with the smallest amplitudes (${\Delta I<0.04}$
mag). The dashed lines show the theoretical {\it PL} relations obtained
assuming that the red giants fill up the Roche lobe. The mass of the red
giant and mass of the companion were both set equal to 1\MS. In our
calculations we used Paczy{\'n}ski's (1971) relation between the mean
radius of the Roche lobe and the separation between components, which with
the use of the Kepler law, leads to the following relation between the
orbital period and mean density of the Roche lobe filling primary
$$P_{\rm orb}=0.12\sqrt{\frac{\bar\rho_\odot}{\bar\rho}}f(q)~~~{\rm [days]}\eqno(1)$$
where
$$f(q)\equiv\sqrt{(1+q)(0.38-0.2\log q)^3}$$
and $q$ is the secondary to primary mass ratio. The $f(q)$ dependence is
weak: ${f=3}$ and 2.1 at ${q=1}$ and 0.1, respectively. The values of $W_I$
were calculated from Girardi \etal (2002) evolutionary tracks for the red
giants phase at the LMC composition (${X=0.74}$, ${Z=0.008}$). We used
Kurucz (1998) stellar atmosphere models to convert the evolutionary model
parameters to the $W_I$ values.

\renewcommand{\TableFont}{\tiny}
\MakeTableSepp{
l@{\hspace{3pt}}r@{\hspace{4pt}}c@{\hspace{2pt}}c@{\hspace{1pt}}c@{\hspace{2pt}}c@{\hspace{3pt}}c@{\hspace{3pt}}c@{\hspace{3pt}}c@{\hspace{4pt}}c@{\hspace{3pt}}c@{\hspace{1pt}}c}
{12.5cm}{Candidates for ellipsoidal variables from the field LMC\_SC1 (first 70 objects)}
{\hline
\noalign{\vskip3pt}
\multicolumn{1}{c}{Star ID} & \multicolumn{1}{c}{Star} & \multicolumn{1}{c}{RA} & \multicolumn{1}{c}{DEC} & $P$ & $T_0$ & $I$ & $V$ & $B$ & $\Delta I_1$ & $\Delta I_2$ & Remarks\\
& \multicolumn{1}{c}{No.} & [J2000] & [J2000] & [days] & [HJD] & [mag] & [mag] & [mag] & [mag] & [mag] & \\
\noalign{\vskip3pt}
\hline
\noalign{\vskip3pt}
OGLE053404.71$-$703351.1 & 184022 & 5:34:04.71 & $-$70:33:51.1 & 401.2  & 2450241.1 & 15.050 & 16.751 &   --   & 0.049 & 0.035 & E \\
OGLE053450.55$-$703332.7 & 266566 & 5:34:50.55 & $-$70:33:32.7 & 203.5  & 2450397.8 & 15.873 & 17.362 & 18.802 & 0.041 & 0.040 &  \\
OGLE053359.09$-$703214.4 & 184031 & 5:33:59.09 & $-$70:32:14.4 & 237.9  & 2450353.7 & 15.332 & 17.021 & 18.586 & 0.038 & 0.027 &  \\
OGLE053308.28$-$703134.8 &  94879 & 5:33:08.28 & $-$70:31:34.8 & 276.6  & 2450318.2 & 15.588 & 17.182 & 19.674 & 0.026 & 0.022 &  \\
OGLE053441.00$-$703117.0 & 266590 & 5:34:41.00 & $-$70:31:17.0 & 286.5  & 2450305.1 & 15.449 & 17.107 &   --   & 0.042 & 0.031 &  \\
OGLE053453.25$-$703114.8 & 266550 & 5:34:53.25 & $-$70:31:14.8 & 395.7  & 2450153.5 & 14.906 & 16.716 & 18.280 & 0.063 & 0.054 & LMC\_SC16 \\
OGLE053337.07$-$703111.7 &  94832 & 5:33:37.07 & $-$70:31:11.7 & 315.5  & 2450366.2 & 14.740 & 16.761 & 18.479 & 0.121 & 0.094 &  \\
OGLE053446.60$-$703005.2 & 271034 & 5:34:46.60 & $-$70:30:05.2 & 275.9  & 2450291.0 & 15.981 & 17.428 & 18.814 & 0.026 & 0.017 & E \\
OGLE053358.74$-$702945.1 & 187738 & 5:33:58.74 & $-$70:29:45.1 & 147.8  & 2450420.4 & 16.044 & 17.277 & 18.371 & 0.019 & 0.015 &  \\
OGLE053434.16$-$702909.1 & 271103 & 5:34:34.16 & $-$70:29:09.1 & 108.2  & 2450402.3 & 16.467 & 17.763 & 18.988 & 0.027 & 0.025 &  \\
OGLE053241.45$-$702908.5 &   4310 & 5:32:41.45 & $-$70:29:08.5 & 225.7  & 2450315.7 & 15.655 & 17.268 & 18.860 & 0.040 & 0.028 & U \\
OGLE053419.84$-$702904.2 & 187743 & 5:34:19.84 & $-$70:29:04.2 & 185.7  & 2450446.1 & 15.778 & 17.281 & 18.737 & 0.085 & 0.075 &  \\
OGLE053244.29$-$702710.7 &   9238 & 5:32:44.29 & $-$70:27:10.7 & 107.1  & 2450383.8 & 16.228 & 17.355 & 18.341 & 0.023 & 0.020 &  \\
OGLE053500.22$-$702643.4 & 275298 & 5:35:00.22 & $-$70:26:43.4 & 286.5  & 2450233.4 & 15.215 & 16.705 & 18.138 & 0.035 & 0.016 & U \\
OGLE053421.62$-$702640.5 & 191935 & 5:34:21.62 & $-$70:26:40.5 & 396.6  & 2450353.3 & 14.739 & 16.550 & 18.275 & 0.095 & 0.068 &  \\
OGLE053347.67$-$702551.4 & 103381 & 5:33:47.67 & $-$70:25:51.4 & 194.6  & 2450275.2 & 15.860 & 17.028 & 18.091 & 0.025 & 0.019 &  \\
OGLE053422.80$-$702355.3 & 192008 & 5:34:22.80 & $-$70:23:55.3 & 165.9  & 2450408.8 & 15.983 & 17.384 & 19.666 & 0.052 & 0.045 &  \\
OGLE053401.95$-$702149.7 & 196753 & 5:34:01.95 & $-$70:21:49.7 & 309.7  & 2450192.5 & 16.086 & 17.711 & 19.278 & 0.051 & 0.033 & E \\
OGLE053452.53$-$702148.3 & 280177 & 5:34:52.53 & $-$70:21:48.3 & \hspace{6.8pt}79.09 & 2450407.5 & 16.720 & 17.963 & 19.171 & 0.025 & 0.021 &  \\
OGLE053505.76$-$702135.0 & 280041 & 5:35:05.76 & $-$70:21:35.0 & 274.5  & 2450411.5 & 15.227 & 16.915 & 20.484 & 0.040 & 0.031 & LMC\_SC16 \\
OGLE053443.62$-$702116.1 & 280043 & 5:34:43.62 & $-$70:21:16.1 & 476.4  & 2450445.1 & 14.793 & 16.196 & 17.404 & 0.020 & 0.017 &  \\
OGLE053349.45$-$701921.5 & 201766 & 5:33:49.45 & $-$70:19:21.5 & 116.7  & 2450373.7 & 16.687 & 17.978 & 19.190 & 0.028 & 0.024 &  \\
OGLE053356.79$-$701919.6 & 201696 & 5:33:56.79 & $-$70:19:19.6 & 205.6  & 2450306.5 & 16.305 & 17.689 & 18.946 & 0.035 & 0.023 & E \\
OGLE053308.07$-$701705.0 & 113390 & 5:33:08.07 & $-$70:17:05.0 & 114.9  & 2450404.6 & 16.072 & 17.512 & 18.894 & 0.053 & 0.042 &  \\
OGLE053303.01$-$701651.7 &  19730 & 5:33:03.01 & $-$70:16:51.7 & \hspace{6.8pt}73.18 & 2450418.5 & 17.078 & 18.288 & 20.213 & 0.039 & 0.027 &  \\
OGLE053506.92$-$701632.1 & 290414 & 5:35:06.92 & $-$70:16:32.1 & 857.9  & 2449618.7 & 14.917 & 16.738 &   --   & 0.049 & 0.013 & E,LMC\_SC16 \\
OGLE053348.74$-$701608.3 & 206947 & 5:33:48.74 & $-$70:16:08.3 & \hspace{6.8pt}89.95 & 2450385.1 & 16.704 & 18.064 & 19.552 & 0.067 & 0.067 &  \\
OGLE053301.61$-$701538.5 &  25473 & 5:33:01.61 & $-$70:15:38.5 & \hspace{6.8pt}68.17 & 2450404.2 & 16.677 & 17.913 & 19.109 & 0.082 & 0.073 &  \\
OGLE053255.66$-$701312.2 &  25439 & 5:32:55.66 & $-$70:13:12.2 & 195.8  & 2450320.3 & 15.993 & 17.405 & 18.828 & 0.017 & 0.012 &  \\
OGLE053406.23$-$701149.8 & 212174 & 5:34:06.23 & $-$70:11:49.8 & 102.1  & 2450419.3 & 16.300 & 17.773 & 19.173 & 0.039 & 0.033 &  \\
OGLE053233.48$-$701021.7 &  31634 & 5:32:33.48 & $-$70:10:21.7 & \hspace{6.8pt}96.36 & 2450430.6 & 16.121 & 17.462 & 18.796 & 0.078 & 0.068 & LMC\_SC2 \\
OGLE053318.71$-$700941.6 & 124851 & 5:33:18.71 & $-$70:09:41.6 & 179.3  & 2450323.8 & 16.406 & 17.769 & 19.166 & 0.032 & 0.027 &  \\
OGLE053355.08$-$700823.8 & 217550 & 5:33:55.08 & $-$70:08:23.8 & 163.6  & 2450381.0 & 16.483 & 17.995 & 19.418 & 0.023 & 0.020 &  \\
OGLE053510.14$-$700755.0 & 301214 & 5:35:10.14 & $-$70:07:55.0 & 496.7  & 2450113.0 & 14.886 &   --   &   --   & 0.031 & 0.021 & LMC\_SC16 \\
OGLE053327.14$-$700751.6 & 130673 & 5:33:27.14 & $-$70:07:51.6 & 343.2  & 2450175.0 & 15.239 & 16.835 & 18.339 & 0.039 & 0.028 & E \\
OGLE053415.75$-$700718.7 & 217491 & 5:34:15.75 & $-$70:07:18.7 & 234.5  & 2450255.8 & 15.115 & 16.985 & 18.729 & 0.170 & 0.121 &  \\
OGLE053423.77$-$700706.8 & 217589 & 5:34:23.77 & $-$70:07:06.8 & \hspace{6.8pt}80.20 & 2450373.2 & 16.759 & 17.971 & 19.083 & 0.067 & 0.053 &  \\
OGLE053226.48$-$700604.7 &  44635 & 5:32:26.48 & $-$70:06:04.7 & 454.1  & 2450428.2 & 15.047 & 16.898 & 18.688 & 0.068 & 0.052 & E,LMC\_SC2 \\
OGLE053422.56$-$700546.6 & 223126 & 5:34:22.56 & $-$70:05:46.6 & 290.6  & 2450286.1 & 15.603 & 17.352 & 19.106 & 0.029 & 0.023 & E \\
OGLE053357.47$-$700529.1 & 223087 & 5:33:57.47 & $-$70:05:29.1 & 395.1  & 2450150.8 & 14.798 & 16.771 & 18.687 & 0.097 & 0.064 &  \\
OGLE053238.26$-$700433.6 &  44721 & 5:32:38.26 & $-$70:04:33.6 & 165.9  & 2450393.4 & 15.348 & 16.733 & 18.110 & 0.026 & 0.018 & LMC\_SC2 \\
OGLE053233.26$-$700433.5 &  44720 & 5:32:33.26 & $-$70:04:33.5 & 229.4  & 2450293.9 & 15.600 & 17.194 & 18.721 & 0.038 & 0.032 & LMC\_SC2 \\
OGLE053419.00$-$700421.3 & 223158 & 5:34:19.00 & $-$70:04:21.3 & 135.7  & 2450439.0 & 16.027 & 17.664 & 19.315 & 0.063 & 0.042 &  \\
OGLE053324.13$-$700241.2 & 137222 & 5:33:24.13 & $-$70:02:41.2 & \hspace{6.8pt}23.48 & 2450430.7 & 16.924 & 17.542 & 17.823 & 0.070 & 0.051 &  \\
OGLE053504.54$-$700239.1 & 307018 & 5:35:04.54 & $-$70:02:39.1 & \hspace{6.8pt}47.22 & 2450408.0 & 16.744 & 18.019 & 18.990 & 0.070 & 0.043 & LMC\_SC16 \\
OGLE053456.76$-$700217.0 & 312758 & 5:34:56.76 & $-$70:02:17.0 & 261.1  & 2450219.9 & 15.066 & 17.133 &   --   & 0.097 & 0.076 & LMC\_SC16 \\
OGLE053420.26$-$700213.8 & 229502 & 5:34:20.26 & $-$70:02:13.8 & 404.2  & 2450082.3 & 15.530 & 16.856 & 17.855 & 0.022 & 0.019 &  \\
OGLE053433.25$-$700132.0 & 312798 & 5:34:33.25 & $-$70:01:32.0 & 128.4  & 2450402.7 & 15.924 & 17.406 & 18.714 & 0.028 & 0.021 &  \\
OGLE053358.34$-$700054.4 & 229486 & 5:33:58.34 & $-$70:00:54.4 & 310.5  & 2450222.7 & 15.109 & 16.803 & 18.798 & 0.094 & 0.075 &  \\
OGLE053444.35$-$700031.4 & 312922 & 5:34:44.35 & $-$70:00:31.4 & 122.8  & 2450356.3 & 16.419 & 17.788 & 19.169 & 0.026 & 0.021 &  \\
OGLE053258.12$-$700020.5 &  51937 & 5:32:58.12 & $-$70:00:20.5 & 110.6  & 2450448.5 & 16.435 & 17.826 &   --   & 0.071 & 0.067 &  \\
OGLE053422.95$-$695814.1 & 235827 & 5:34:22.95 & $-$69:58:14.1 & 468.0  & 2450436.3 & 15.371 & 16.902 & 18.527 & 0.038 & 0.031 &  \\
OGLE053459.27$-$695755.1 & 318682 & 5:34:59.27 & $-$69:57:55.1 & \hspace{6.8pt}78.74 & 2450428.3 & 15.408 & 16.686 & 17.740 & 0.168 & 0.091 & LMC\_SC16 \\
OGLE053319.66$-$695714.6 & 150929 & 5:33:19.66 & $-$69:57:14.6 & 220.1  & 2450323.3 & 14.776 & 15.868 & 16.560 & 0.033 & 0.026 &  \\
OGLE053241.26$-$695648.0 &  59891 & 5:32:41.26 & $-$69:56:48.0 & \hspace{6.8pt}15.67 & 2450435.2 & 17.730 & 18.878 & 19.872 & 0.130 & 0.112 &  \\
OGLE053358.19$-$695643.6 & 235791 & 5:33:58.19 & $-$69:56:43.6 & 204.6  & 2450430.5 & 15.029 & 16.601 & 18.081 & 0.117 & 0.113 &  \\
OGLE053305.72$-$695639.8 &  59340 & 5:33:05.72 & $-$69:56:39.8 & 481.6  & 2450266.0 & 15.522 & 17.143 & 18.816 & 0.038 & 0.033 &  \\
OGLE053413.75$-$695636.1 & 235794 & 5:34:13.75 & $-$69:56:36.1 & 311.7  & 2450140.9 & 14.842 & 16.541 & 18.541 & 0.025 & 0.018 &  \\
OGLE053248.55$-$695634.6 &  59343 & 5:32:48.55 & $-$69:56:34.6 & 391.6  & 2450367.2 & 15.564 & 17.043 & 18.404 & 0.018 & 0.015 &  \\
OGLE053438.78$-$695634.1 & 318659 & 5:34:38.78 & $-$69:56:34.1 & 384.7  & 2450296.1 & 14.699 & 16.692 & 18.415 & 0.049 & 0.045 &  \\
OGLE053305.00$-$695623.7 &  59348 & 5:33:05.00 & $-$69:56:23.7 & 125.8  & 2450370.2 & 16.179 & 17.629 & 18.947 & 0.047 & 0.039 &  \\
OGLE053351.92$-$695604.8 & 235799 & 5:33:51.92 & $-$69:56:04.8 & 233.5  & 2450432.1 & 14.981 & 16.197 & 17.370 & 0.018 & 0.005 &  \\
OGLE053302.21$-$695537.8 &  59366 & 5:33:02.21 & $-$69:55:37.8 & 208.2  & 2450332.4 & 15.500 & 17.292 & 18.891 & 0.052 & 0.040 &  \\
OGLE053245.16$-$695524.7 &  66515 & 5:32:45.16 & $-$69:55:24.7 & \hspace{6.8pt}67.61 & 2450404.2 & 15.905 & 17.236 & 18.493 & 0.118 & 0.067 &  \\
OGLE053418.79$-$695523.1 & 242700 & 5:34:18.79 & $-$69:55:23.1 & \hspace{6.8pt}53.25 & 2450406.6 & 16.615 & 17.858 & 19.609 & 0.095 & 0.083 &  \\
OGLE053500.93$-$695355.2 & 325028 & 5:35:00.93 & $-$69:53:55.2 & 216.1  & 2450360.0 & 15.295 & 16.988 & 18.633 & 0.123 & 0.091 & LMC\_SC16 \\
OGLE053409.59$-$695311.0 & 242787 & 5:34:09.59 & $-$69:53:11.0 & \hspace{6.8pt}39.41 & 2450426.7 & 16.602 & 17.764 & 18.799 & 0.139 & 0.110 &  \\
OGLE053309.47$-$695238.1 & 158043 & 5:33:09.47 & $-$69:52:38.1 & 229.1  & 2450224.0 & 15.390 & 16.823 & 18.267 & 0.045 & 0.016 &  \\
OGLE053309.75$-$694957.5 & 164389 & 5:33:09.75 & $-$69:49:57.5 & 232.7  & 2450252.5 & 15.685 & 17.237 & 18.802 & 0.030 & 0.025 &  \\
OGLE053304.04$-$694908.9 &  73016 & 5:33:04.04 & $-$69:49:08.9 & 371.0  & 2450353.5 & 15.209 & 16.985 & 18.644 & 0.047 & 0.023 &  \\
\noalign{\vskip2pt}
\hline
\noalign{\vskip1pt}
\multicolumn{11}{l}{Remarks: U -- uncertain, E -- eccentric orbit, LMC\_SC? -- the same stars in the field LMC\_SC?}
}

It is clearly seen that dispersion of the sequence~E depends on the
amplitude of variability. The amplitude of ellipsoidal variation is a
function of the third power of the Roche-lobe filling factor (Hall 1990),
therefore the variables with the largest amplitudes should be close to the
limiting lines. Periods of variables with smaller amplitudes of ellipsoidal
modulation may significantly depart from the theoretical limits, because
giants do not entirely fill the Roche lobe.

The dispersion of data seen in the $\log P$--$W_I$ diagrams may be caused
in part by the spread in the red giant masses. For instance, the models
with ${M{=}1.5~\MS}$ have the $W_I$ values brighter by some 0.3~mag than
the models with ${M=1~\MS}$ of the same $P_{\rm orb}$. Another contribution
to the dispersion may arise from the the expected spread in the secondary
mass, through the $f(q)$ factor in Eq.~(1).

Different depths of minima, shown by the vast majority of our ellipsoidal
candidates, can be explained by higer order effects of the tidal
distortion. In addition, there is a number of variables in which different
height of the light curve maxima is visible. This, so called O'Connel
effect (Davidge and Milone 1984), is usually explained as a result of spots
in the stellar surface. In some cases maximum luminosity of these objects
clearly changes with time, which is probably caused by the star-spot
changes. Some unequal maxima may be a~result of eccentric orbit of the
system (see Section~6).

\Section{Ellipsoidal Variability and Long Secondary Periods}
Long Secondary Period (LSP) is one of the last unexplained types of stellar
variability. At least 30\% of the variable red giants show periods which
are an order of magnitude longer than typical pulsation period of a
semi-regular variable. Wood \etal (1999) noticed that the LSPs form a
separate sequence in the period--luminosity space (sequence ``D'').

Many hypotheses have been proposed to explain the LSP. Radial and
non-radial pulsation, periodic dust ejection, rotation of the spotted star
and binarity have been suggested as the origin of the LSPs. Recently Wood,
Olivier and Kawaler (2004) ruled out most of these propositions. They
argued that the most likely explanation of the LSP is a low degree ${\rm
g^+}$ mode of pulsation trapped in the outer layers above the convective
envelope.

\begin{figure}[!tb]
\centerline{\includegraphics{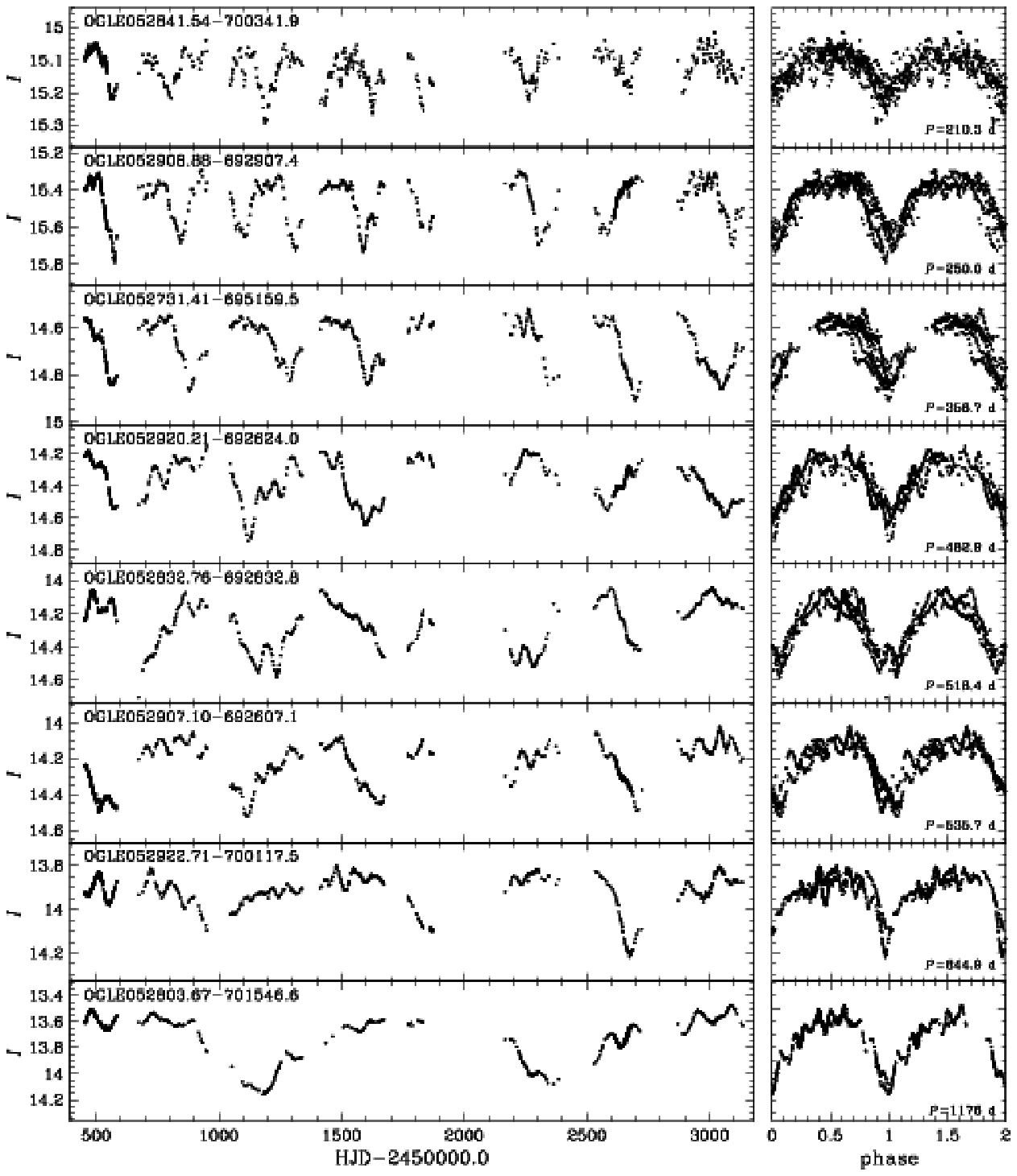}} 
\FigCap{Light curves of the exemplary LSP red giants in the LMC.} 
\end{figure} 
We selected from OGLE database giants with dominant period corresponding to
the LSP, and then we visually inspected their light curves. We removed from
the list all uncertain and doubtful variables, and left about 1600 stars in
which amplitudes of the LSP are significantly larger than amplitudes of
other variations. Several typical light curves folded with the long periods
are shown in Fig.~4.

\begin{figure}[htb]
\vspace{-1cm}
\centerline{\includegraphics[width=12cm]{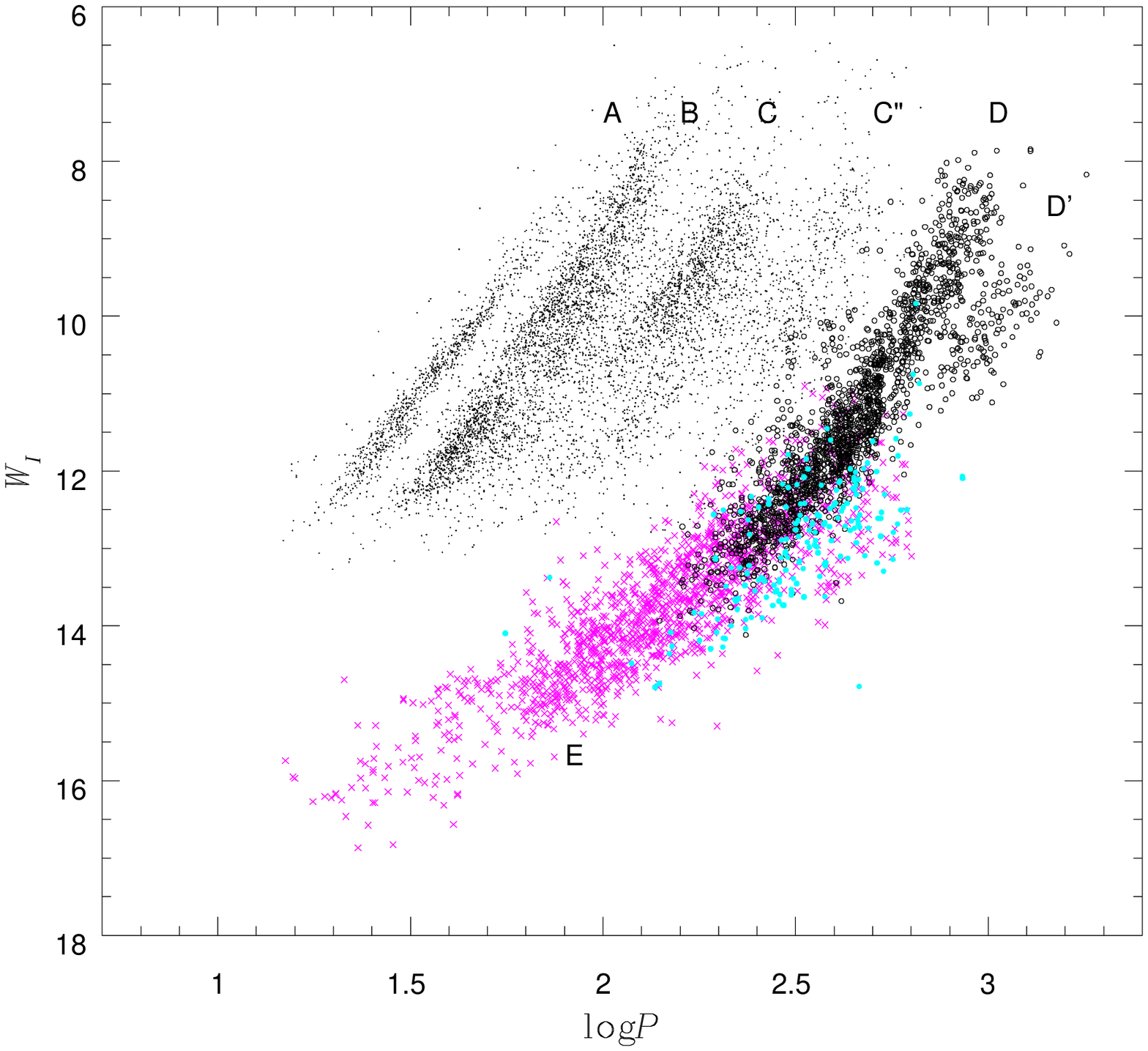}} 
\vspace{-4.7cm}
\FigCap{Period--$W_I$ diagram for variable red giants in the LMC. Magenta
crosses mark candidates for ellipsodal variables, cyan dots show
ellipsoidal variables with eccentric orbits, empty circles indicate LSPs,
and small dots show other red giants.}
\end{figure} 
The period--$W_I$ diagram for the dominant periods of these objects is
shown in Fig.~5. Selected LSPs are marked with the empty circles, the
magenta crosses indicate our candidates for the ellipsoidal variables. It
is worth noticing that apart from the ${P{-}L}$ sequence D presented for
the first time by Wood \etal (1999), Fig.~5 reveals additional sequence
(marked by D$'$) of stars with longer periods.

It is not surprising that the ellipsoidal variability leads to a rather
narrow band in the $\log P$--$W_I$ diagram. Along the red giant branch
stellar radius, $R$ increases by more than two orders of magnitude. The
luminosity increase is dominated by the radius increase and the period,
which cannot be much longer than $P_{\rm orb}$ given in Eq.~(1), is
determined primarily by the value of $R^{1.5}$.

The striking feature clearly seen in Fig.~5 is that the sequence of
ellipsoidal variables partly overlaps with the sequence~D and is a
continuation of this sequence toward fainter stars. It might suggest binary
origin of the LSP phenomenon in the red giants.

\begin{figure}[tb]
\vspace{-0.4cm}
\centerline{\includegraphics{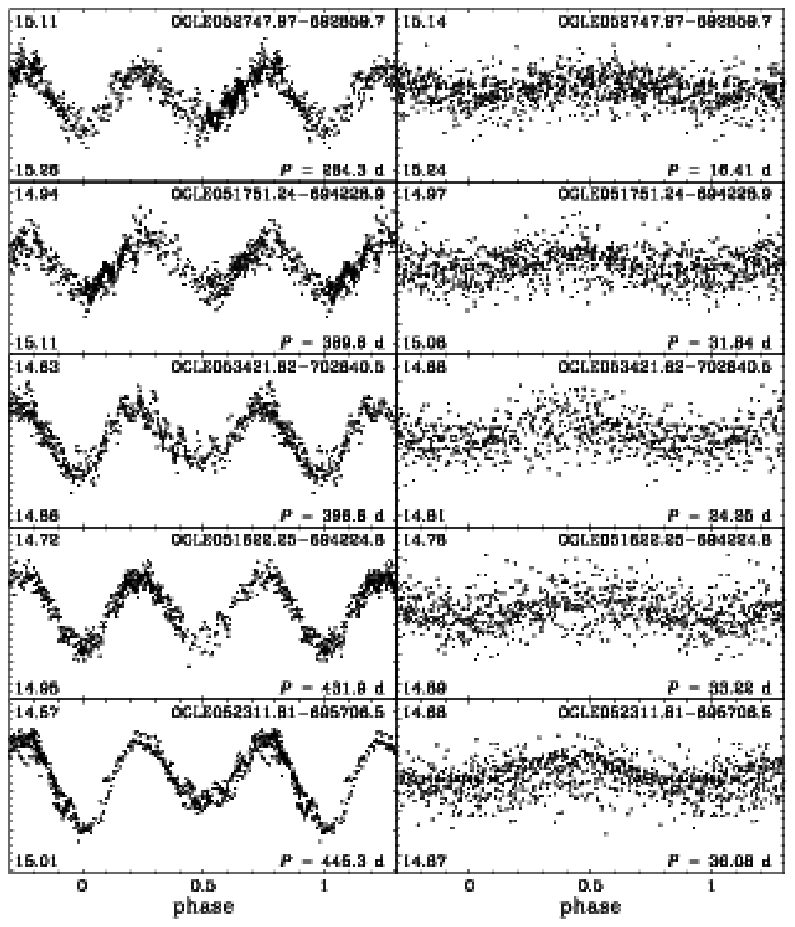}} 
\FigCap{Light curves of the ellipsoidal candidates exhibiting
simultaneously OSARG-type variability. {\it Left column} shows folded
ellipsoidal light curves, and {\it right column} presents OSARG light
curves after subtracting the dominant variability.}
\end{figure} 
Moreover, there is no doubt that some of the LSP cases may be explained by
ellipsoidal variability. Fig.~6 presents exemplary light curves of stars
where the characteristic OSARG variability is superimposed on the evident
ellipsoidal light curves. We found about 300 such objects, mostly among
brighter stars. The shorter, small amplitude variability corresponds to
OSARG's period--luminosity sequences (Soszy{\'n}ski \etal 2004).

However, there are several serious differences between ellipsoidal
variables and typical LSP giants. First, typical LSP light curve is not
double humped, like the ellipsoidal light modulation (\cf Fig.~1 and
Fig.~4). Second, the typical period -- amplitude relation for the LSP
variables is different than for the ellipsoidal giants. The amplitudes of
LSP variability are positively correlated with the brightness of the star,
while in ellipsoidal variables the amplitudes are connected only with the
dispersion of the {\it PL} relation (Fig.~3). Third, Wood, Olivier and Kawaler
(2004) noticed that the binary origin of the LSP is difficult to accept on
statistical ground. LSP occurs in one third of the AGB stars, while the
ellipsoidal variability relates to less than 1\% of the analysed giants.

The $\log P$--$W_I$ correlation for the LSP is very significant and cannot
be ignored in any explanation of phenomenon. If the stars are forced to
rotate at nearly break up velocity, then we expect a $P_{\rm
rot}(\bar\rho)$ relation, the same as that given in Eq.~(1), but with
${f=1.2}$. Any asymmetry about the rotation axis could produce light
variation with the $P_{\rm rot}$. However, the implied equatorial
velocities of rotation seems unacceptably large. For instance at ${\log
P\approx2.5}$ we obtain the value of about 30~km/s.

The explanation favored by Wood, Olivier and Kawaler (2004) (an excitation
of a g-mode trapped above the convective envelope) may also explain the
observed correlation. However, properties of such modes and chances of
their driving remain to be studied.

\Section{Ellipsoidal Systems with Eccentric Orbits}
About 10\% of candidates for ellipsoidal variables show evident deviations
from symmetric sinusoidal light curves. For a sample of 165 objects the
shapes of the light curves are completely different than typical sine-wave
light curves of the ellipsoidal variables. Examples of these light curves
are presented in Fig.~7. One should notice a large variety of the light
curves.

\begin{figure}[tb]
\centerline{\includegraphics{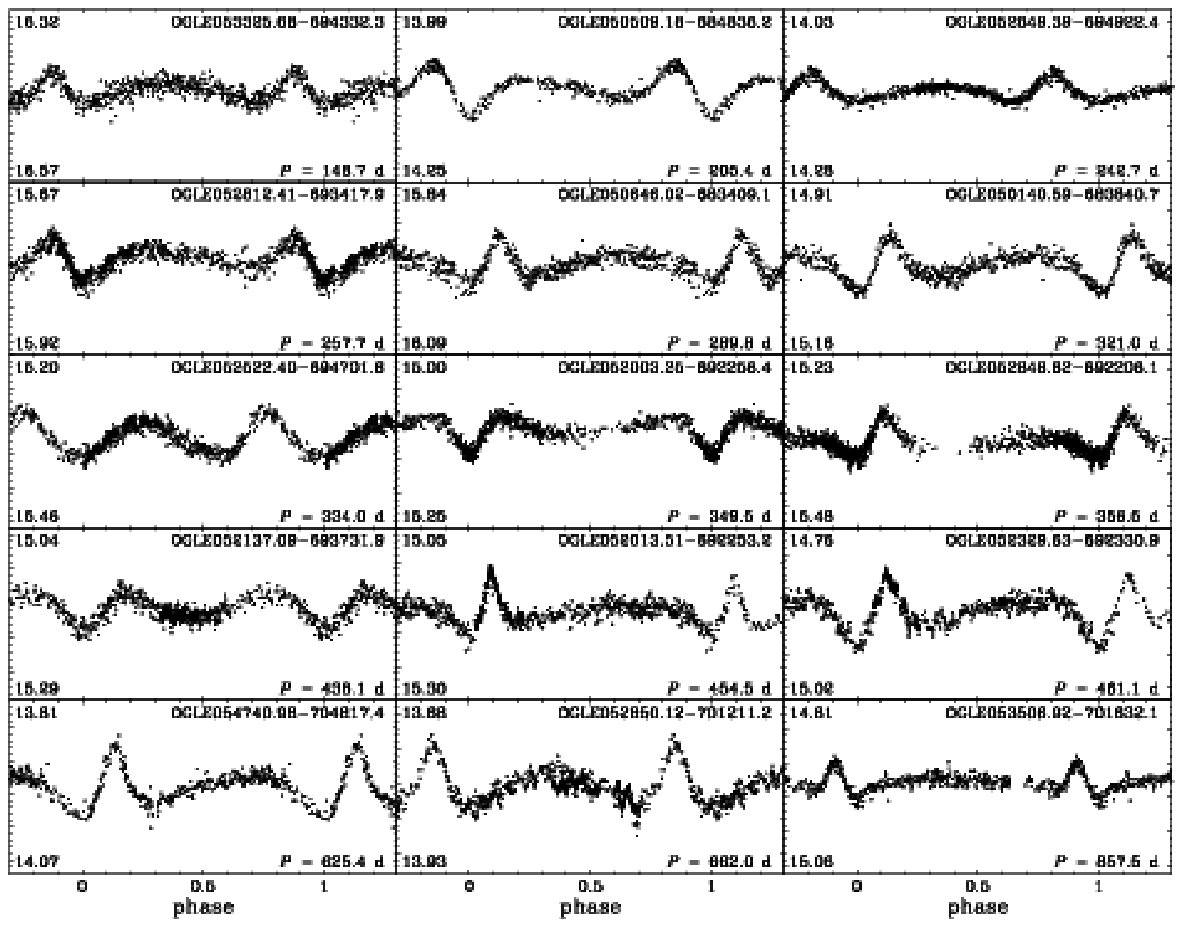}} 
\FigCap{Light curves of the exemplary ellipsoidal red giants with 
eccentric orbits.}
\end{figure}

An explanation of such variables was proposed by B.~Paczy{\'n}ski (2004,
private communication) who suggested that these objects may be ellipsoidal
systems with significant eccentricity of the orbits. A similar periodic
light curve, although with the period of about 4 days, is shown by the
well-studied star $\alpha$~Vir (Spica). Photometric and spectroscopic
measurements showed that it is an ellipsoidal binary system with
significant eccentricity of the orbit, equal to 0.14. Variable distance of
the binary components changes tidal interaction of stars during the orbital
movement, which affects the shape of the ellipsoidal modulation. Moreover,
the $\alpha$~Vir light curve is changing in the time scale of several dozen
years (\cf the light curves obtained by Shobbrook
\etal 1969 and Sterken, Jerzykiewicz and Manfroid 1986), what is an effect
of apsidal line rotation. Large variety of the light curves observed in the
LMC can be explained by different longitude of the line of apsides.

\begin{figure}[tb]
\vspace{-1cm}
\centerline{\includegraphics[width=14cm]{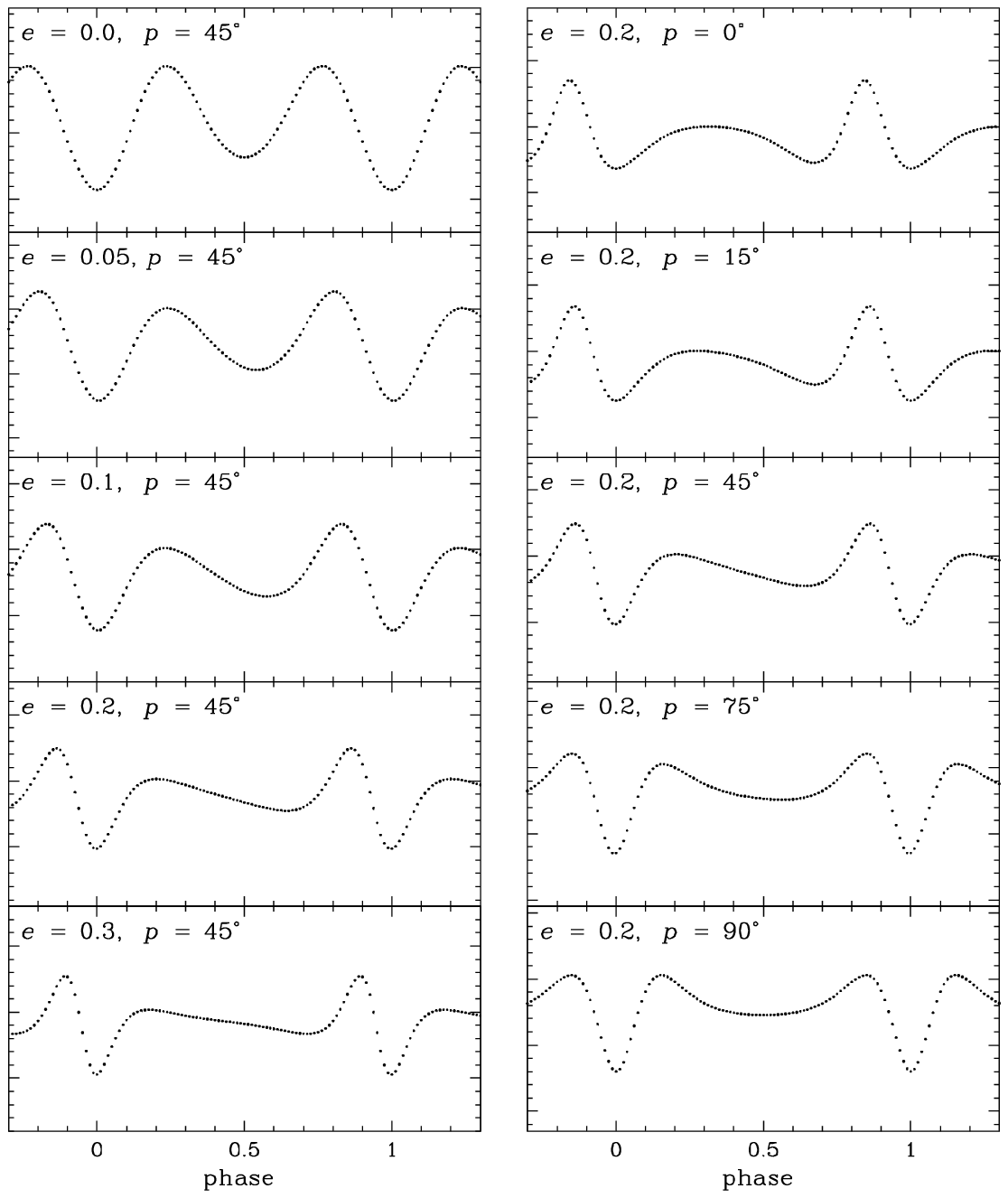}} 
\vspace{-8.0cm}
\FigCap{Model light curves of the ellipsoidal red giants with eccentric
orbits. In the {\it left column} light curves with increasing eccentricity
($e$) are showed, the {\it right column} presents a series of light curves
produced by binary systems with the eccentricity ${e=0.2}$, and a different
periastron length~($p$).}
\end{figure}

To make sure that the variables are binary systems with eccentric orbits,
we modeled such light curves using program {\sc
Nightfall}\footnote{{\it http://www.lsw.uni-heidelberg.de/users/rwichman/Nightfall.html}}
by R.~Wichmann. A series of models are presented in Fig.~8. We modeled an
equal-mass binary system with one of the components entirely filling the
Roche lobe. The inclination angle is equal to 60\arcd\!. In the left column
the light curves with increasing eccentricity~($e$) are showed, while in
the right column we present light curves produced by binaries with the same
eccentricity (${e=0.2}$), but with a different periastron length ($p$). One
can note clear similarity of the observed and calculated light curves.

The models confirm that the larger eccentricity of the orbit, the more
asymmetric light curve is. The variable showing one of the most asymmetric
light curve -- OGLE052013.51--692253.2 -- corresponds to a model with
${e=0.4}$. However, precise modeling of the binary systems will be possible
after obtaining the radial velocity curves.

Verbunt and Phinney (1995) estimated that the circularization time scale of
the semi-detached binaries containing a giant is of the order of several
thousand years. This estimation is almost independent of the mass of the
companions. Thus, it is very unlikely that the close binaries maintained
their initial eccentricity. It is more probable that the eccentricity of
the orbits was induced by the rapid mass transfer or interaction of the
third companion.

Mechanisms of the orbit circularization induced by the tidal effects were
studied by Zahn (1977, 1989) and Tassoul and Tassoul (1992). Such processes
are especially effective for the red giants, due to the turbulent viscosity
associated to the convective envelopes. Studies of the spectroscopic binary
systems containing red giants showed that there exist limiting period below
which circularization takes place. Mayor and Mermilliod (1984) analyzed the
orbital parameters of 17 spectroscopic binaries with a red giant, and found
that orbits are circularized at periods ${P<150}$ days, however there are
many exceptions to this rule.

Obviously, the circularization time scale depends not only on the
semi-major axis of the orbit (and consequently orbital period), but also on
the radius of the giant star (related to the luminosity of the object). One
can notice this effect in Fig.~5, where the period--luminosity relation of
the candidates for eccentric ellipsoidal variables is marked by cyan
dots. The magnitudes of objects are clearly correlated with the
periods: the brighter stars, the longer orbital period of the
system. Moreover, eccentric ellipsoidal variables have typically longer
periods than non-eccentric ellipsoidal systems.

\Section{Summary}
In this paper we showed that the ellipsoidal red giants constitute a
numerous and homogeneous sample. This class of variables is difficult to
detect because of their small amplitudes and long periods. Only long-term
observing surveys, obtaining good quality photometry, can trace this type
of variability.

Ellipsoidal variables form the {\it PL} sequence what is a projection of
the radius--luminosity relationship for the red giants. We noticed that
{\it PL} sequence of the ellipsoidal red giants is a continuation of the
sequence of LSP variables, what may suggest a connection of the LSP and the
binarity. Some ellipsoidal variables of the longer periods exhibit
small-amplitude variability, corresponding to OSARGs.

About 10\% of the ellipsoidal variables show characteristic asymmetric
light curves, probably caused by close binary systems with eccentric
orbits. This sample may become an important test of tidal circularization
theories.

\Acknow{We would like to thank Prof. Bohdan Paczy{\'n}ski, whose
suggestions enabled us to explain the mystery of ellipsoidal variables with
eccentric orbits. The paper was partly supported by the Polish KBN grant
2P03D02124 to A.~Udalski. Partial support to the OGLE project was provided
with the NSF grant AST-0204908 and NASA grant NAG5-12212 to
B.~Paczy\'nski.}

\end{document}